\documentclass{aa}  
\usepackage{graphicx}
\begin{document}
\title{Confirmation of the Planet Hypothesis for the Long-period Radial 
Velocity Variations of $\beta$ Geminorum}


   \author{A. P. Hatzes\inst{1}
          \and
   W. D. Cochran\inst{2}
          \and
   M. Endl\inst{2}
          \and
   E. W. Guenther\inst{1}
          \and
          S. H. Saar\inst{3}
	 \and
	G.A.H. Walker\inst{4}
	 \and
	S. Yang\inst{5} 
	\and
	M. Hartmann\inst{1}
	 \and
	M. Esposito\inst{1}
         \and
	D. B. Paulson\inst{6}
             }

   \offprints{
    Artie Hatzes, \email{artie@tls-tautenburg.de}\\$*$~
    Based in part on observations obtained at the 
    2-m-Alfred Jensch Telescope at the Th\"uringer
    Landessternwarte Tautenburg and the Harlan J. Smith
   2.7m telescope of McDonald Observatory
  \\}

     \institute{Th\"uringer Landessternwarte Tautenburg,
                Sternwarte 5, D-07778 Tautenburg, Germany
		\and 
	McDonald Observatory, The University of Texas at Austin,
    Austin, TX 78712, USA
              \and
Harvard-Smithsonian Center for Astrophysics, 60 Garden
Street, Cambridge, MA 02138
\and
1234 Hewlett Place, Victoria, BC, V8S 4P7, Canada
\and
Department of Physics and Astronomy, University of Victoria,
Victoria, BC, Canada, V8W 3P6
\and
Planetary Systems Branch, Code 693, NASA Goddard Space Flight Center, Greenbelt, MD 20771, USA 
}

   \date{Received; accepted}

 
  \abstract{
   Our aim is to confirm the nature of the long period radial velocity
measurements for $\beta$ Gem first found by Hatzes \& Cochran (1993).
We present precise stellar radial velocity measurements
for the K giant star $\beta$ Gem spanning over 25 years.
An examination of the Ca II K emission,
spectral line shapes from high resolution data
($R$ = 210,000), and Hipparcos photometry  was also made to discern
the true nature of the long period radial velocity variations.
The radial velocity  data show that the
long period, low amplitude radial velocity  variations
found by Hatzes \& Cochran (1993) are
long-lived and coherent. Furthermore, the Ca II K emission,
spectral line bisectors, and Hipparcos photometry  show no significant
variations of these quantities with the radial velocity period. 
An orbital solution
assuming a stellar mass of 1.7 $M_\odot$ yields  a period, $P$ = 589.6 days,
a minimum mass
of 2.3  $M_{Jupiter}$, and a semi-major axis, $a$ = 1.6 AU. The orbit is
nearly circular ($e$ = 0.02).
   The  data presented here confirm the
planetary companion hypothesis suggested by Hatzes \& Cochran (1993).
$\beta$ Gem is the sixth intermediate mass star shown to host
a sub-stellar companion and suggests that planet-formation around stars
much more massive than the sun may common.

\titlerunning{A Planetary Companion to $\beta$ Gem}
}
\maketitle

%

\section{Introduction}

\begin{figure*}
\resizebox{\hsize}{!}{\includegraphics{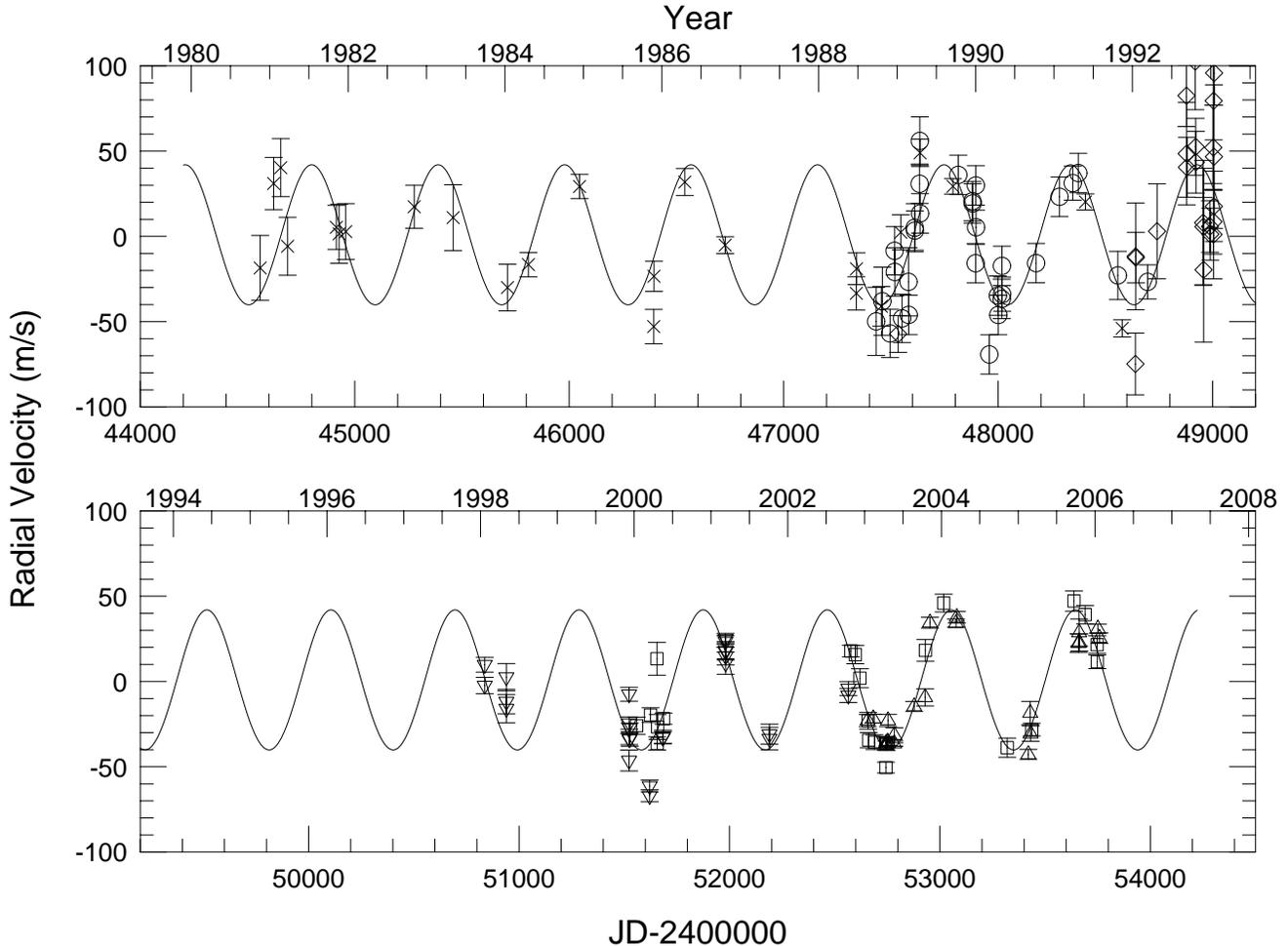}}
\caption{Radial velocity measurements for $\beta$ Gem from the 6 data sets:
CFHT (crosses), DAO (diamonds), McD-2.1m (circles), 
McD-cs21 (inverted triangles), McD-MOPS (squares), and TOPS (triangles).
}
\label{orbit}
\end{figure*}

Long-period, low amplitude radial
velocity (RV) variations were reported in three K giant stars, including $\beta$ Gem ($=$ Pollux 
$=$ HR 2990 $=$ HD 62509) by 
Hatzes \& Cochran 1993 (hereafter HC93). One hypothesis for the RV variations was the 
presence of planetary companions. In the case of $\beta$ Gem an orbital 
solution yielded a minimum mass $m$ sin $i$ = 2.9 $M_{Jupiter}$
(assuming a stellar mass of $M$ = 2.8 $M_\odot$), semi-major
axis, $a$ = 1.9 AU and eccentricity, $e$ = 0.12. Hatzes \& Cochran noted that ``...it would seem 
that planetary companions around K giants have been detected." 
 However, since RV variations of comparable periods were also
found for $\alpha$ Boo and $\alpha$ Tau, HC93 were cautious in the
interpretation: ``...it seems improbable that all three would have
companions with similar masses and periods unless planet formation
around the progenitors to K giants was an ubiquitous 
phenomenon." Indeed, up until that time extrasolar planet RV searches 
were yielding no detections, even around main-sequence stars, 
so it seemed odd that K giant stars 
would produce an abundance of sub-stellar companions. 
This, and the fact that the 
expected rotation periods of K giant stars were  comparable to the detected
long period RV variations made rotational modulation by stellar surface
structure a viable alternative.

Subsequent work by  Larson et al. (1993b) 
confirmed the long period RV variations of $\beta$ Gem  with a revised 
period of 585 days. That work also showed that the Ca II $\lambda$8662 
equivalent width varied at the few percent level with the same period as the RV
variations. This seemed to support  rotational modulation  by surface structure
as a cause of the RV variations,
although the Ca II equivalent width variations were marginal.
The false alarm probability for the signal was about 1\% and 
Larson et al commented: ``However, because of the weakness of the
signal, $K$ $=$ 0.583 $\pm$ 0.19 m{\AA}, this signal needs confirmation."

Since the discovery of RV variations in $\beta$ Gem, planetary companions
have been established in several K giant stars. The planetary companion
to $\iota$ Dra (Frink et al. 2002) was largely accepted because of the
eccentric orbit, a shape in the RV curve that is difficult to produce with
rotational modulation. Companions to HD 47536 (Setiawan et al. 2003), HD 11977
(Setiawan et al. 2005), and HD 13189 (Hatzes et al. 2005)
were established by the absence of Ca II H \& K emission and the
lack of  spectral
line bisector variations with the same period as the RV.

Over a decade since their discovery the nature of
the  long period variations in $\beta$ Gem is  still unknown. For
these reasons we continued to monitor this star  with
precise stellar radial velocity measurements. These new measurements
along with previous ones span over 25 years in time. 
In Section \S3 
we show that the RV variations
in this star first reported in 1993 are still present with the same
period and amplitude. In Section \S4 we make a careful
analysis of ancillary data that were available to us. These are Ca II
K emission, Hipparcos photometry, and spectral line bisectors 
measured using very high spectral resolution data.
Our analysis demonstrates
that there are no other forms of variability with the same
period as the RV variations. This confirms the planet hypothesis for this
star first reported by HC93.

\section{The Star }

$\beta$ Gem is  a K0 III star at a distance of 
10.3 pcs as measured by Hipparcos. Interferometric measurements
have determined  an angular diameter  of
7.96 $\pm$ 0.09 mas (Nordgren, Sudol, and Mozurkewich (2001) which corresponds
to a radius of 8.8 $\pm$ 0.1 $R_\odot$
The atmospheric parameters for this star have been derived by several
investigators. McWilliam (1990)
measured an effective temperature of 4850 K, a metallicity of 
$[Fe/H]$ $=$ $-$0.07,  and a surface gravity, log $g$ $=$ 2.96.
Gray et al. (2003) derived the same effective temperature, but a lower surface
gravity (log $g$ $=$ 2.52) and higher metallicity ($[Fe/H]$ $=$ $0.08$).
More recently, Allende-Prieto et al. (2004) determined 
$T_{eff}$ = 4666 $\pm$ 95 K, a surface gravity of log $g$ 
$=$ 2.685 $\pm$ 0.09 and a metallicity, $[Fe/H]$ $=$ $0.19$, a value
considerably higher than previous determinations. 

Allende Prieto \& Lambert (1999) 
derived a stellar mass of $M$ $=$ 1.7 $\pm$ 0.4 $M_\odot$, 
a value we shall adopt.  We are aware that the determination of the
stellar mass for giant stars is difficult. This relies not only on 
an accurate measurement of stellar parameters, but also on the reliability
of evolutionary models. Since main sequence stars with spectral types
in the range A--G can all evolve to K giant status, the true mass
of $\beta$ Gem may be outside the range of the nominal error given
by Allende Prieto \& Lambert.
They also derived a stellar radius of $R$ $=$ 8.9 $\pm$ 0.4 $R_\odot$,
a value consistent with interferometric measurements.

\section{The Radial Velocity Data}

Six  independent data sets of high precision radial velocity data were used
for our analysis. Three data sets have  already been published.
Larson et al. (1993b) used data from the Canada France Hawaii Telescope
(CFHT) survey as well as data taken at the Dominion Astrophysical Observatory
(DAO). The DAO data had a RV precision about a factor of 2 worse than the
CFHT measurements as well as the ones were present here.  These
data span the time 1980 -- 1993.
The McDonald Observatory measurements that 
appeared in the discovery paper of HC93 were made at the 2.1m telescope
(McD-2.1m). These data
span the time period 1980 -- 1992. New measurements were obtained at McDonald
Observatory and the Thuringia State Observatory (Th\"uringer Landessternwarte). 
Table 1 lists the journal of
observations which includes the data set, time coverage, the RV technique
employed, and the number of observations. The CFHT and DAO
measurements employed a hydrogen fluoride 
(HF) absorption cell (Campbell \& Walker 1979). 
All other data sets utilized the iodine absorption cell (I2) 
for the wavelength reference.  Table 1 also lists
the rms scatter of the points about
the orbital solution (see below). These standard deviations represent the true measurement
error as well as any intrinsic variability of star on time scales much shorter
than the long period found in the data.

Two of the new  data sets were obtained from McDonald Observatory.
The first set was taken as  part of the Phase III radial velocity program 
of the McDonald Observatory Planet Search  (MOPS) program that used
the ``low resolution'' mode of the 2dcoud{\'e}
cross-dispersed echelle spectrograph
(Tull et al. 1994) at the Harlan J. Smith 2.7m telescope.  
This instrument, when used with a Tektronix 2048$\times$2048
detector,  provides a nominal wavelength coverage of 
3600\,{\AA} -- 1$\mu$m at a resolving power of $R$ ($=\lambda$/$\Delta \lambda$) $=$ 60,000.  
The RV-information from the I$_2$ self-calibrated 
spectra taken during Phase III was obtained using the
$Austral$ RV-code (Endl, K\"urster, \& Els 2000).
The data from the McDonald Phase III  are given in Table 2.
The uncertainties quoted there and in other tables are the ``internal''
errors as represented by the the rms scatter of the individual 
spectral chunks ($\sim$ several hundred) used in the
RV calculation (see Endl et al. 2000). 
We regard these as a lower limit on the actual uncertainties,
since these values do not include the effects of any residual systematic
errors or stellar `jitter' that may be present.   

The second set of McDonald data were taken using the high
resolution mode of the 2dcoud{\'e} cross-dispersed echelle spectrograph
(often referred to as the cs21 focus and referred to as `cs21' in the
tables and figures).
This setup provided a resolving power, $R$ = 210,000 using the same
Tektronix CCD detector, although with much more limited wavelength
coverage (about
800\,{\AA}). These data were primarily taken for an examination of the
spectral line shapes, although some observations were made with the iodine
cell so as to correlate any RV variations with changes in the spectral line shapes.
Table 3 lists the RVs made with the high resolution mode of the
2dcoud{\'e}.


\begin{table}
\begin{center}
\begin{tabular}{ccccc}
Data Set  &  Coverage & Technique & N & $\sigma_{RV}$  \\
  &           &           &   &               (m\,s$^{-1}$) \\
\hline
CFHT                 & 1980.87--1991.87 & HF cell      & 24      & 20.3 \\
DAO                  & 1985.28--1993.05 & HF cell      & 25      & 37.7 \\
McD-2.1m             & 1988.73--1992.20 & Iodine Cell  & 31      & 18.9  \\
McD-MOPS             & 1998.69--2000.03 & Iodine Cell  & 22      & 11.6  \\
McD-cs21             & 2000.03--2004.58 & Iodine Cell  & 11      & 16.7  \\
TLS-TOPS             & 2003.04--2006.06 & Iodine Cell  & 22      & 11.0  \\
\hline
\end{tabular}
\caption{The data sets used in the orbital solution.}
\end{center}
\end{table}

\begin{table}
\begin{center}
\begin{tabular}{lrr}
JD   & RV (m/s)  & $\sigma$ (m/s) \\
\hline
2451558.8477 & $-$25.93 & 5.0  \\
2451624.7461 & $-$19.67 & 4.3 \\
2451655.5820 & 13.34  & 9.6 \\
2451656.5898 & $-$36.34 & 3.9 \\
2451658.5938 & $-$26.64 & 7.3 \\
2451686.6055 & $-$21.78 &  3.3 \\
2452577.0117 & 17.86  & 3.4 \\
2452597.9805 & 15.74  & 5.3 \\
2452619.9531 & 1.93   & 5.5 \\
2452658.8828 & $-$23.03 & 4.9 \\
2452661.7227 & $-$34.38 & 4.5 \\
2452688.8594 & $-$35.57 &  4.3 \\
2452742.5938 & $-$36.13 &  4.0  \\
2452743.5938 & $-$50.45 & 3.2 \\
2452931.0078 & 18.27  & 6.4 \\
2453017.8711 & 45.97  & 5.2 \\
2453320.0078 & $-$38.81 & 5.6 \\
2453436.6523 & $-$28.23 & 3.7 \\
2453636.9766 & 47.20  & 6.0  \\
2453689.9219 & 39.20  & 5.4 \\
2453746.9062 & 21.60  & 5.1 \\
2453747.8125 & 11.56  & 4.0 \\
\hline
\end{tabular}
\caption{Radial Velocity Measurements for $\beta$ Gem from the McDonald 
Phase III program (MOPS).}
\end{center}
\end{table}

\begin{table}
\begin{center}
\begin{tabular}{lrr}
JD   & RV (m/s)  & $\sigma$ (m/s) \\
\hline
2450836.7266  &      3.72   &     6.4 \\
2450939.6445  &     $-$8.38   &     9.0  \\
2451522.9219  &    $-$26.92   &     7.2  \\
2451523.9141  &    $-$29.07   &     4.8  \\
2451525.9219  &    $-$30.54   &     5.0  \\
2451620.7695  &    $-$64.17   &     4.9  \\
2451683.6562  &    $-$32.67   &     5.2  \\
2451980.7500  &     24.25   &     4.3  \\
2451981.6406  &     15.37   &     5.4  \\
2452189.9844  &    $-$32.17   &     8.4  \\
2452565.0000  &     $-$6.17   &     5.6  \\
\hline
\end{tabular}
\caption{Radial Velocity Measurements for $\beta$ Gem taken 
with the high resolution  mode of the
2dcoud{\'e} spectrometer.}
\end{center}
\end{table}

\begin{table}
\begin{center}
\begin{tabular}{lrr}
JD   & RV (m/s)  & $\sigma$ (m/s) \\
\hline
2452656.6562 & $-$22.87 & 3.5 \\
2452683.2109 & $-$21.72 & 2.5 \\
2452744.2891 & $-$36.09 & 3.1 \\
2452746.2969 & $-$37.64 & 2.8 \\
2452751.2891 & $-$35.31 & 2.1 \\
2452752.2930 & $-$35.72 & 2.1 \\
2452753.2969 & $-$22.83 & 3.5 \\
2452782.8359 & $-$36.32 & 2.6 \\
2452783.3359 & $-$31.51 & 4.4 \\
2452877.6172 & $-$14.63 & 2.9 \\
2452929.6055 & $-$9.30  & 5.0 \\
2452952.5391 & 34.42  & 3.2 \\
2453076.2656 & 34.23  & 2.4 \\
2453080.3047 & 37.66  & 3.3 \\
2453420.3789 & $-$42.72 & 2.9 \\
2453429.3594 & $-$18.47 & 6.8 \\
2453431.3477 & $-$30.56 & 4.5 \\
2453658.6523 & 22.47  & 5.5 \\
2453661.6055 & 28.74  & 8.0  \\
2453662.6680 & 22.77  & 5.0 \\
2453750.4609 & 30.94  & 2.6 \\
2453758.3281 & 25.43  & 3.0 \\
\hline \\
\end{tabular}
\caption{Radial Velocity Measurements for $\beta$ Gem from TOPS.}
\end{center}
\end{table}

Finally, observations of $\beta$ Gem  were made as part of the Tautenburg
Observatory Planet Search (TOPS) program. This uses 
the
high resolution coud{\'e} echelle spectrometer  of the Alfred-Jensch 2m
telescope and an iodine absorption cell placed in the optical
path. This is a grism crossed-dispersed  echelle spectrometer that has
a resolving power $R$ ($\lambda$/$\Delta \lambda$) = 67,000 and 
wavelength coverage 4630--7370\,{\AA} when using the so-called ``Visual''
grism. A more detailed description of radial velocity measurements
from the TOPS program can be found in Hatzes et al. (2005). Table 4 lists
RV measurements of $\beta$ Gem from the TOPS program.

The RV measurements for all data sets are shown in Figure~\ref{orbit}.
Each data set had its own velocity offset  that
had to be applied 
so that they would all have the same zero point (see below, the tabulated
values have the offsets applied). There 
is a clear sinusoidal
variation that persists over the entire time span covered by the data.

\begin{figure}[h]
\resizebox{\hsize}{!}{\includegraphics{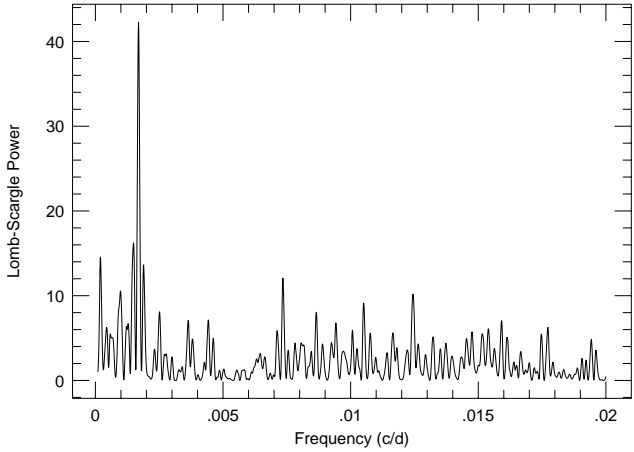}}
\caption{Lomb-Scargle periodogram of the combined RV data sets.
}
\label{periodogram}
\end{figure}

Figure~\ref{periodogram} shows the  Lomb-Scargle
periodogram  (Lomb 1976, Scargle 1982) of all the RV measurements.
These show a dominant peak at a frequency of $\nu$ = 0.0017 c\,d$^{-1}$  (period = 590 d).
The false alarm probability (FAP) of this peak using 
the expressions in Scargle (1982)
is  FAP $\approx$ 10$^{-16}$. With such large Lomb-Scargle power it is pointless
to perform Monte Carlo simulations of the FAP. Nevertheless, the FAP 
was also was determined using the bootstrap randomization technique 
(Murdoch et al. 1993; K\"urster et al. 1997).  The measured RV values were
randomly shuffled keeping the observed times fixed and 
a periodogram for the shuffled data computed. 
The fraction of the random periodograms
having power higher than the 
data periodogram yields 
the false alarm probability that noise would create the detected
signal. As expected, after 2$\times 10^{5}$ ``shuffles'' 
there was no instance
where the random periodogram had more Lomb-Scargle power than the
data. This FAP is indeed very small.

\section{Orbital Solution}

\begin{figure}[h]
\resizebox{\hsize}{!}{\includegraphics{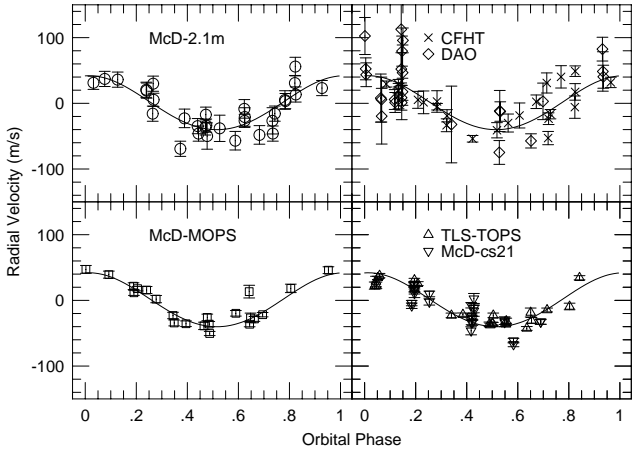}}
\caption{Radial velocity measurements for $\beta$ Gem from the 6 data sets phased
to the orbital period. The symbols are the same as for Fig. 1.
}
\label{phase}
\end{figure}

An orbital solution was calculated using the  combined data sets. This
is shown by the solid line in Fig.~\ref{orbit}. The velocity zero point for each
data set was allowed to be a free parameter. This  was varied 
until the best fit in a least-squares sense was obtained.
These individual zero points in 
the velocity were subtracted from each data set before plotting in 
Fig.~\ref{orbit}. 
The orbital parameters are listed in Table 5.  The rms scatter listed for the 
orbital solution is from the combined data sets. 
Figure~\ref{phase}
shows the RV measurements of each data set
phased to the orbital period.
We should note that a significant fraction of the 
the rms scatter may be due to intrinsic
variability. K giants are known to exhibit stellar oscillations with 
periods of 0.25 -- 10 days and amplitudes of 10--100 m\,s$^{-1}$
(Hatzes \& Cochran 1994ab, 1995, Kim et al. 2006).

\begin{table}
\begin{center}
\begin{tabular}{lr}
Parameter  & Value  \\
\hline
Period [days]  &   589.64 $\pm$ 0.81\\
T$_{periastron}$ [JD] & 2447739.02  $\pm$  4.5 \\
$K$ [m\,s$^{-1}$] &  41.0 $\pm$ 1.6 \\
$e$               & 0.02 $\pm$ 0.03 \\
$\omega$ [deg]    &  354.58 $\pm$ 95.65\\
$f(m)$   [solar masses] & (4.21 $\pm$ 0.48) $\times 10^{-9}$ \\
$m$ sin $i$ [$M_{Jupiter}$]       & 2.30 $\pm$ 0.45\\
$a$ [AU] & 1.64 $\pm$ 0.27\\
rms [m\,s$^{-1}$] & 20.6 (17.1) \\
\hline
\end{tabular}
\caption{Orbital parameters for the companion to $\beta$ Gem. The
rms scatter in parenthesis is without  the DAO measurements.}
\end{center}
\end{table}

\section{The Nature of the RV Variations}

The fact that the RV variations seem to be long-lived and coherent
for over 25 years strongly argues that they are indeed due to a sub-stellar
companion. However, $\beta$ Gem is a giant star and we know little about the 
nature of possible surface structure on these type of stars or how long-lived
they might be. An exotic form of long-period stellar oscillation
can also not be excluded. Furthermore, the weak Ca II $\lambda$8662 variations found by Larson 
et al.  (1993b) compels us to be cautious about the interpretation of the RV 
variations. To confirm that a sub-stellar companion is indeed responsible
for the RV variations we examined the Ca II K emission, the spectral line 
shapes, and the Hipparcos photometry to see if any of these correlated with
the RV variations.

\subsection{Ca II emission}

Larson et al. (1993b) found  variations in the equivalent width of
Ca II 8662 that showed a long term trend on a timescale greater than 12 years.
This was fit with a quadratic polynomial plus sinusoid that had
the same period as the RV variations. As stated earlier the
periodogram of the Ca II equivalent width measurements
  after subtracting the long term showed only marginal power at the
RV period. Larson et al. (1993a) showed that their Ca II 8662 measurements
gave results consistent with the Mt. Wilson S-index measurements.
The wavelength coverage of the McDonald Phase III data included the Ca II
K line, a feature traditionally used for measurement of
stellar chromospheric variability.
Paulson et al. (2002) defined  an 
S-index that did not include the 
Ca {\sc ii} H line  since it was contaminated by a 
strong Balmer H$\epsilon$ feature.
The mean  McDonald  S-index (SMcD, and not to be confused with the 
Mt. Wilson S-index) for
$\beta$ Gem is 0.118 $\pm$ 0.005  (see \S2.3 in Paulson et al. 2002 for a 
detailed description of how the McDonald 
S-index for the Ca II K line core emission is obtained). This low value of
the S-index is comparable to the inactive dwarf star $\tau$ Ceti 
(SMcD $=$ 0.0166 $\pm$ 0.008). However, because of calibration issues
it is probably not appropriate to compare the McDonald S-index between
giants and dwarfs.   On the other hand, support for the inactivity of 
$\beta$ Gem comes from the X-ray flux for this star which is about a
factor of 50 less than that of the solar value (Rutten et al. 1991). 

\begin{figure}[h]
\resizebox{\hsize}{!}{\includegraphics{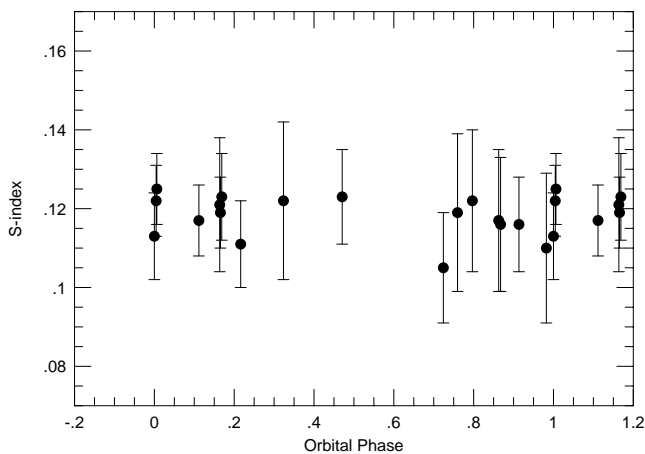}}
\caption{McDonald S-index measurements for $\beta$ Gem phased to the
590-day orbital period. Points are repeated for the second 
cycle.
}
\label{caiiphase}
\end{figure}

\begin{figure}[h]
\resizebox{\hsize}{!}{\includegraphics{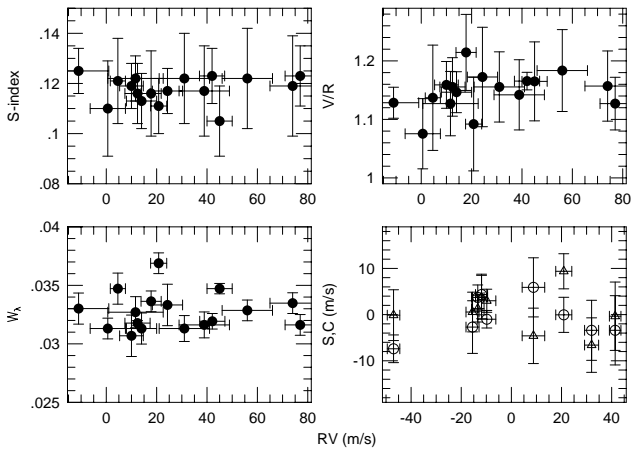}}
\caption{Correlation of the radial velocity with S-index (top left panel),
ratio of emission peaks in Ca II, $V/R$ (top right panel),
equivalent width of core Ca II emission (bottom left panel), and velocity
span, S (circles) and curvature, C (triangles)  of the
spectral line bisector (lower right panel).
}
\label{correlations}
\end{figure}

For the purpose of confirming the nature of the RV variations we are not
so much interested in the mean activity level of the star, but rather the
{\it variability} of this level. The McDonald S-index which is calculated in
a consistent way from the same dataset is appropriate for such an
investigation.  Figure~\ref{caiiphase} shows the 
the McDonald S-index measurements phased to orbital period. There
are no obvious sinusoidal variations. The upper left  panel of
Figure~\ref{correlations} shows the S-index measurements versus
the RV measurement. The two quantities are uncorrelated having a
correlation coefficient, $r$ $=$ 0.07 and a probability, $p_{uncor}$ $=$ 0.78 
that they are uncorrelated. 

K giant stars are known to exhibit variations in the Ca II core  emission
peaks, the ratio often denoted by $V/R$. For example, Arcturus shows variations
in $V/R$ from 0.80 to 1.05 (Gray 1980). The nature of these variations
are not known. One possibility is variable mass loss (Chiu et al. 1977).
Regardless of the cause, any variations in $V/R$ that are correlated with
the RV variations would cast doubt on the planet hypothesis. We measured the 
$V/R$ ratio for the Ca II K line and these are plotted versus the
RV measurement in the upper right panel of Figure~\ref{correlations}.
Again there  is  no obvious correlation between the two quantities 
($r$ = 0.27,  $p_{uncor}$ = 0.27).

Finally, the total equivalent widths of the two $K$ core emission peaks
with respect to the flux level of the core of the line 
on either side of the peaks were also measured. The variations of these
with RV are shown in the lower left  panel of Figure~\ref{correlations}.
This quantity is also not correlated with the RV variations
($r$ $=$ $-$0.03, $p_{uncor}$ $=$ 0.9).
Our analysis of the Ca II data fails to support the hypothesis
that the RV variations are due to magnetic  (chromospheric) activity.

\subsection{Spectral Line Bisector Variations}

The analysis of the shapes of spectral lines via line bisectors has proved
to be an effective technique for confirming the planet hypothesis for RV variations.
A lack of spectral line bisectors provided the final confirmation of the planet
hypothesis to 51 Peg (Hatzes, Cochran, \& Bakker 1998).
Constant spectral line bisectors have established that sub-stellar companions 
were responsible for the RV variations in the K giants 
HD 47536 (Setiawan et al. 2003), HD 11977 (Setiawan et al. 2005), and HD 13189
(Hatzes et al. 2005).  

To investigate whether $\beta$ Gem exhibits line shape variations, observations
were taken with the high resolution mode of the 2dcoud{\'e} spectrograph
($R$ = 210,000)
on 10 different nights. The phase sampling of the data was
good between phase 0 and 0.6, but  with a large phase gap between 0.6--1.0. In spite of the large
phase gap the phase coverage was sufficient to show any  possible sinusoidal variations.
Four to five exposures each with 
signal-to-noise levels of greater than 300 were taken.
Spectral line bisectors
were computed for 11 strong, unblended spectral features. 
For our bisector measurements we chose the spectral lines 
Fe\,I~$\lambda$5379.6,
Fe\,I~$\lambda$5543.2, Fe\,I~$\lambda$5637.4, Fe\,I~$\lambda$5731.8, 
Fe\,I~$\lambda$5934.7, Fe\,I~$\lambda$6141.7,
Fe\,I~$\lambda$6151.6, Fe\,I~$\lambda$6252.6,  Fe\,I~$\lambda$6254.2,  
Ca\,I~$\lambda$6499.6, and Fe\,I~$\lambda$6750.1.
Two bisector quantities were measured: the velocity span which is the velocity
difference between two endpoints of the bisector and the curvature which is
the difference of the velocity span of the upper half of the bisector minus the
lower half. We examined both quantities because it is possible for a star to show
variations in one quantity but not the other.
For our span measurements we chose flux levels of 0.40 and 0.85
of the continuum and 0.6 for the curvature measurement. 
These avoided the cores and wings of the spectral lines where the
error of the bisector measurements are large. The  average  velocity span and
curvature were computed for each spectral line and for each observation. After
subtracting the mean value of the bisector span (curvature)  for each 
line all bisectors  quantities 
were averaged together to produce the  mean for
a given night. Thus approximately 50--60 individual bisector
measurements (4--5 individual observations and 11 spectral lines) go into the
computation of each mean value at a given orbital phase.

\begin{figure}[h]
\resizebox{\hsize}{!}{\includegraphics{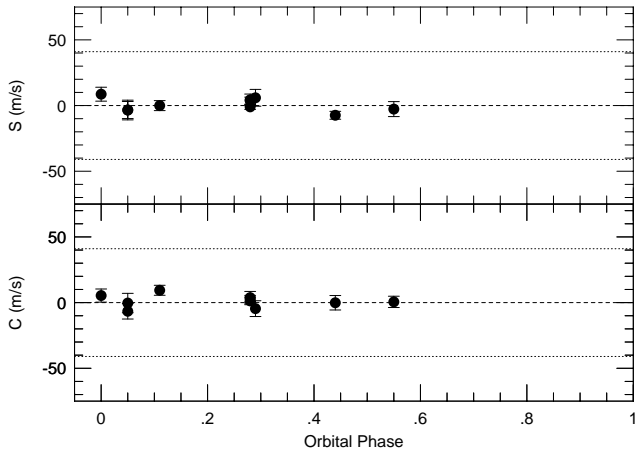}}
\caption{The mean velocity span (top) and curvature (bottom) of the
spectral line bisectors of $\beta$ Gem as a function of orbital phase
(590-day RV period). Dotted lines mark the maximum extent of the
RV variations and the dashed line is the zero level.
}
\label{bisectors}
\end{figure}

\begin{figure}[h]
\resizebox{\hsize}{!}{\includegraphics{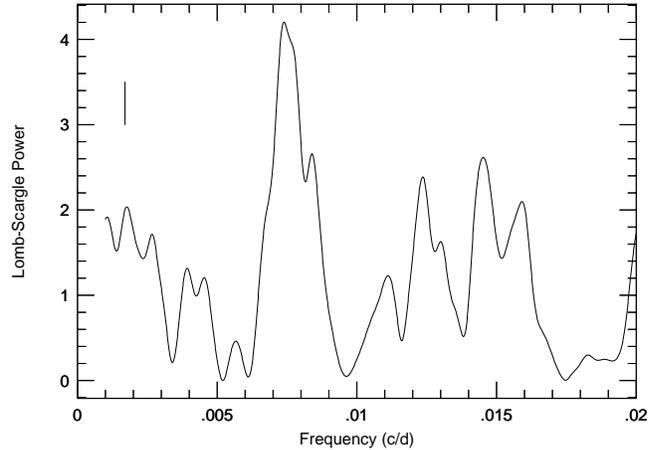}}
\caption{The Lomb-Scargle periodogram of the Hipparcos photometry.
The vertical line marks the orbital frequency.
}
\label{ftphot}
\end{figure}

\begin{figure}[h]
\resizebox{\hsize}{!}{\includegraphics{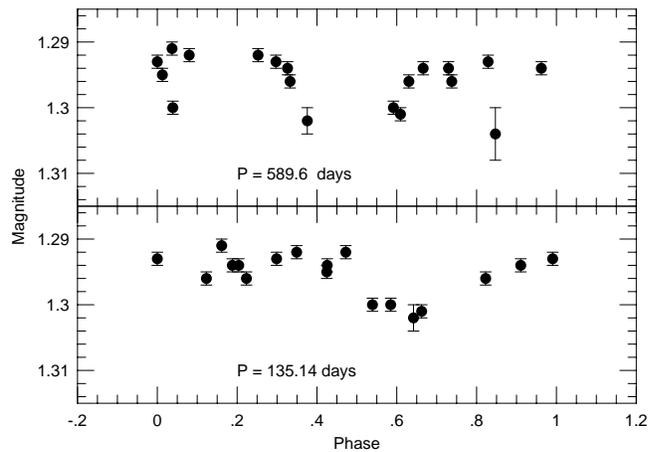}}
\caption{The Hipparcos photometry for $\beta$ Gem phased to the 590-d orbital period (top)
and the best-fit 135-d period (bottom). The measurement with the large
error was excluded in the lower panel. }
\label{phot1}
\end{figure}

\begin{figure}[h]
\resizebox{\hsize}{!}{\includegraphics{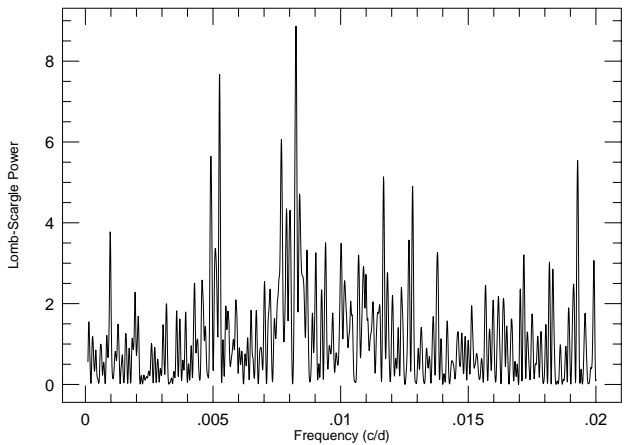}}
\caption{The Lomb-Scargle Periodogram of the residual RV measurements
after subtracting the contribution of the orbital motion due to the
companion.
}
\label{ftres}
\end{figure}

Figure~\ref{bisectors} shows the resulting bisector velocity span and curvature  measurements
phased to the period found in the RV data. 
The error bars represent the standard error of the mean
(standard deviation of measurements used for each average, 
divided by the square root of the number of measurements).
There are no convincing sinusoidal RV variations in either
the velocity span or curvature.
A least squares sine fit to the data
yields an amplitude of 2.5 $\pm$ 2.8 m\,s$^{-1}$ for span variations 
and 0.40  $\pm$ 2.8  m\,s$^{-1}$ for any curvature variations in the spectral line shapes.
The lower right panel of Figure~\ref{correlations}  shows the 
bisector velocity span and curvature values
versus the RV measurements. The correlation coefficient is 0.10 with  a probability of 0.79  for the
data not being correlated. Our analysis of the spectral line shapes also does not support
rotational modulation by surface features or pulsations as a cause for the radial velocity 
variations.

\subsection{Hipparcos Photometry}

The Hipparcos satellite made precise photometric measurements
for $\beta$~Gem that were contemporaneous with the RV measurements of our data set.
Figure~\ref{ftphot} shows the Lomb-Scargle periodogram of the 
Hipparcos photometry excluding one measurement with large error (four times
the average error). The top panel of 
Figure~\ref{phot1} shows this photometry phased to the 590-d orbital
period. Although there is considerable scatter in the data there are no
obvious sinusoidal variations.
A sine fit to the photometry using the orbital period yields
an amplitude of 3.03 $\pm$ 3.12 mmag for photometric  variations with the RV period.

A least squares sine-fit was made to the Hipparcos photometry again
excluding the one point with large error.
This yielded a best fit period of 135 days,  consistent with the highest
peak of the periodogram.  The lower panel of
Figure~\ref{phot1} shows the photometry phased to the 135-day period.
If this period is indeed present then it most likely represents the
rotation period of the star. More importantly,
the Hipparcos photometry does not support rotational modulation 
as the cause of the RV variations. 

We analyzed the residual RV variations after subtracting the
Keplerian motion to see if we could detect any evidence of the 
135-day photometric period in the RV measurements. Figure~\ref{ftres}
shows the Lomb-Scargle periodogram of the residual RV variations excluding
the lower precision DAO measurements.
The highest peak is at a frequency corresponding to a period of
121 days.  (The second highest peak is for a period of 190 days).
Although this is near the period of the best-fit sine 
wave to the photometric data we do not consider it as significant. The
false alarm probability is  0.017 determined
after 10,000 shuffles of the bootstrap randomization technique.
(We consider a FAP $<$ 0.001 to be a significant periodic signal.)
However, both the Hipparcos photometry and the residual RV measurements
show some evidence that the rotation period of $\beta$ Gem
may be  $\sim$ 130 days.

\section{Discussion}

Our analysis of the radial velocity measurements for $\beta$~Gem   show
that the long period variations found by HC93 and confirmed by Larson
et al. (1993a) are long-lived and coherent. These RV variations have not
changed in period, 
amplitude, or  phase over the past 25 years. A careful examination of the
Ca II K emission, spectral line shapes, and Hipparcos photometry
reveals no convincing variation with the 590-day RV period. If the RV variations
were due to stellar surface structure or stellar oscillations then it
is difficult to
reconcile the RV variations with a lack of spectral or photometric 
variability. Of course, we cannot entirely exclude that an exotic form of stellar 
oscillations could cause the RV variations. For example, toroidal modes
have all of their atmospheric motion in the horizontal direction. These
would produce no photometric or Ca II emission variations. However,
these modes can produce line profile variations if the star is viewed
from an intermediate inclination (Osaki 1986). Our bisector measurements
exclude this possibility.
The most likely and logical explanation for the RV variations is that
they are indeed due to a planetary companion with minimum mass of 2.3 $M_\odot$
at an orbital distance of 1.6 AU. These data confirm the planet hypothesis
for the long period RV variations first proposed by HC93.

If the 135-d period found in the Hipparcos photometry indeed
represents the rotational period, then this can be used to estimate
the stellar inclination. HC93 measured a projected rotational 
velocity for $\beta$ Gem of 1.6 $\pm$ 0.9 km\,s$^{-1}$. A radius
of 8.8 $R_\odot$ yields an equatorial rotational velocity of
3.3  km\,s$^{-1}$. This 
yields sin $i$ $=$ 0.48 $\pm$ 0.3. Assuming an alignment of rotational
and orbital axes results in a true companion mass of 
2.9 -- 12.8 $M_{Jupiter}$.

We can check if the possible photometric variations that are detected
could  result from rotation in spite of the low activity level of this
star by comparing it with models.  Given the period and velocity amplitude
of the RV variations radial pulsations would produce a change in stellar
radius of about 10\%. This can be excluded by the lack of large photometric
variations.
Non-radial pulsations are ruled
out by lack of bisector variations, and significant numbers of
starspots would be surprising on such an inactive star (e.g.,
$F_X/F_X(\odot) \approx 0.02$; Rutten et al 1991).  The average solar
spot coverage is $f_S \sim$ 0.001 ($\langle R_M \rangle = 50, f_S
\approx 2.2\times 10^{-5} R_M$; Cox 2000).  If $f_S$ scales with
$F_X$, then for $\beta$ Gem $f_S \sim 2 \times 10^{-5}$, far too small
to yield the photometric variability seen by Hipparcos ($\sigma_V
\approx 0.0033$).

A more likely possibility is microvariability due to the stellar
granulation -- specifically, due to the finite number of (variable)
convective granules on the stellar surface.  Ludwig (2006) estimates
that the fractional flux RMS $\sigma_F/F$ due to granulation is given
by (combining their equations 56 and 59):

$\sigma_F/F = 0.4 N^{-0.5} (\delta I_{\rm RMS}/I) f(a)$ \\
where $N$ is the number of granules on the stellar surface, $\delta
I_{\rm RMS}/I$ is the fractional RMS intensity variation due to a
granule at $\mu=1$, and $f(a)$ is a slowly varying function of the
linear limb-darkening coefficient $a$.  Freytag et al. (2002) use 3-D
hydrodynamic  models to show that the size of a typical granule scales
with the pressure scale height as

$x_{\rm gran}/R_\ast \approx 0.0025 (T_\ast/T\odot)(R_\ast/R_\odot)
(M_\ast/M_\odot)^{-1}$.

Thus, for $\beta$ Gem $x_{\rm gran}/R_\ast \approx 0.015$, implying $N
\sim 8900$ on the visible stellar surface.  Then, if we adopt the solar
value of  $\delta I_{\rm RMS}/I$ = 0.18, and take $\epsilon$ = 0.8,
$f(a) \approx 1.3$, and $\sigma_F/F \sim 0.001$.  While this a factor
of $\approx$3 smaller than $\sigma_V$, we feel the two are sufficiently
close (given the many approximations involved) to suggest it is quite
probable that the  photometric variation is due to granulation-induced
inhomogeneities combined with rotation.

$\beta$ Gem with an estimated mass of 1.7 $M_\odot$
is the sixth star of intermediate mass known to host an extrasolar
planet. Table 6 lists those stars in the mass range 1.8 -- 5  $M_\odot$ 
known to host giant planets ($\iota$ Dra has
an estimated mass of 1.05 $M_\odot$). 

Although the number of intermediate mass
stars hosting giant planets is small, these already show some interesting
characteristics. First, all would qualify as
``super planets" having   masses much greater than 
a few  Jupiter masses.  Possibly, more massive stars have more massive
protoplanetary disks   which could result in more massive planets.
Second, the semi-major axes for most are around 2 AU. This might raise
concerns that we are seeing some other phenomena such as rotation and not
evidence for planetary companions; however, we believe this is not the
case for two reasons: 1) For these stars the 
RV variations were not accompanied by other forms of variations which
excluded rotational modulation or pulsations as a cause. 2) The derived
orbital eccentricities span a wide range ($e$ = 0.01 -- 0.40). 
If the RV variations
were due to stellar rotation or pulsations, 
then we would not expect similar shapes
in the RV curves and not the wide variety that is provided by 
Keplerian motion.   Furthermore, not all K giant stars show long period
RV variations. In a sample of 62 K giants D\"ollinger et al. (2006) found
evidence for long period RV variations in at most 15\% of  stars.

Most of the stars in Table 6 have low metallicities.
Rice \& Armitage (2005) have argued that stars with lower metallicities
take longer to form planets. These would have had little time to migrate
before the disks were dispersed leaving the giant planets near the snow-line
of $\approx$ 2--4 AU. However, $\beta$ Gem and $\gamma$ Cep have metallicities
considerably higher than the solar value. Alternatively, radiation pressure
from the more massive and thus hotter star may have dispersed the disks
before the giant planet had time to migrate.  However, it is dangerous to 
draw conclusions based on such a small sample.
The sample of exoplanets around 
intermediate stars  must be
increased by at least an order of magnitude before we can discern
the true distribution of semi-major axes and companion masses. Further
discoveries of giant planets around intermediate stars may hold important
clues for planet formation.

\begin{table*}
\begin{center}
\begin{tabular}{cccccccc}
Star & $M_{star}$  &  $R_{star}$ & M$_{planet}$$\times$ sin $i$ & $a$ & $P$ &  $e$ & [Fe/H]   \\
\hline
HD 11977$^1$       & 1.9   & 10.2 & 6.5  & 1.9  & 1420  & 0.40 & $-$0.14  \\
HD 47536$^2$       & 2.0   & 21.3 & 5.0  & 2.0  & 712   & 0.20 & $-$0.61    \\
HD 13189$^3$       & 3.5   & -    & 14   & 1.8  & 471   & 0.27 & $-$0.59 \\
HD 104985$^4$      & 1.6   & 11   & 6.3  & 0.78 & 198   & 0.03 &  $-$0.35 \\
$\gamma$ Cep$^5$   & 1.6   & 4.7  & 1.7  & 2.3  & 920   & 0.12 & $+$0.18 \\
$\beta$ Gem    & 1.7   & 8.4  & 2.3  & 2.4  & 590   & 0.02 &  $+$0.19 \\
\hline
\end{tabular}
\caption{Sub-stellar Companions to Intermediate Mass Stars
($^1$ Setiawan et al. 2005, $^2$ Setiawan et al. 2003,
$^3$ Hatzes et al. 2005, $^4$Sato et al. 2003, $^5$Hatzes et al. 2003)}
\end{center}
\end{table*}
-----------------------------------------------------------------------

\begin{acknowledgements}
We thank Carlos Allende-Prieto for useful discussions.
WDC and ME acknowledge the support of NASA grants NNG04G141G and NNG05G107G.
SHS was supported by NASA Origins of Solar Systems grant NNG04GL54G.
GAHW is supported by the Natural Sciences and Engineering Research Council 
of Canada.  DBP is currently a NASA Postdoctoral fellow. The NASA Postdoctoral
Program is administered by the Oak Ridge Associated Universities.
This research has made use of the SIMBAD data base operated
at CDS, Strasbourg, France.

\end{acknowledgements}

\end{document}